\documentclass[]{aa}
\usepackage{natbib}
\usepackage{psfig}
\usepackage{txfonts}
\def\EBV{\mbox{E$_{\rm B-V}$}}

\def\HH{\mbox{H$_2$}}
         
\def\nH2{{\rm n}({\rm H}_2)}
\def\NH2{{\rm N}({\rm H}_2)}
\def\pccc{~{\rm cm}^{-3}} \def\pcc{~{\rm cm}^{-2}}

\def\Tstar#1 {\mbox{${\rm T}_{\rm #1}^*$}}
\def\Tsub#1 {\mbox{$T_{\rm #1}$}}
\def\TK  {\Tsub K }

\def\Texc {\Tsub exc }

\def\degr{$^{\rm o}$}
\def\p{\mbox{$^+$}}

\def\hcop{\mbox{{HCO\p}}} 
\def\hocp{\mbox{{HOC\p}}}

\def\cch{\mbox{C$_2$H}}
  
\def\hhco{\mbox{H$_2$CO}}
\def\h13cop{\mbox{{H$^{13}$CO\p}}}

\def\c3h2{\mbox{C$_3$H$_2$}}
 
 \def\R0{R$_0$}

\def\ddeg{{}^\circ\kern-.1em}

\def\kms{\mbox{km\,s$^{-1}$}}
\def\ps{\mbox{s$^{-1}$}}
\def\bll{BL Lac}

\def\E#1 {$10^{#1}$}
\def\E#1 {E{#1}}
\def\P#1,{$\nH2\TK~=~#1\times~10^4\pccc$~K}
\def\ec#1,#2,#3,{#1\,(#2)\E{#3}}

\def\zoph{$\zeta$ Oph}

\def\H3{\mbox{H$_3$}}

\def\ammon{\mbox{N\H3} }

\title{The abundance of \hocp\ In diffuse clouds}
\author{H. Liszt\inst{1}\ and R. Lucas\inst{2}\ and J. H. Black\inst{3} }
\institute{National Radio Astronomy Observatory,
           520 Edgemont Road,
           Charlottesville, VA,
           USA 22903-2475
\and       Institut de Radioastronomie Millim\'etrique,
           300 Rue de la Piscine,
           F-38406 Saint Martin d'H\`eres,
           France
\and       Chalmers Centre for Astrophysics and Space Science, 
           Onsala Space Observatory,
           Chalmers University of Technology, 
           43992 Onsala, Sweden}

\begin{document}
\date{received \today}
\offprints{H. S. Liszt}
\mail{hliszt@nrao.edu}
\abstract{
We used the Plateau de Bure Interferometer to search for $\lambda$3mm 
absorption lines of \hocp\ from local diffuse and translucent clouds 
occulting compact extragalactic mm-wave continuum sources.  We detected
\hocp\ in three directions with column densities only 70-120 times below
that of the \hcop\ isomer, a factor 5-50 higher than typically
found in dense dark gas but comparable to recent observations of
dense photon-dominated regions.  The observed amounts of \hocp\ 
N(\hocp)/N(\HH) $ = 3-6\times 10^{-11} $ can be made in {\bf quiescent} 
diffuse gas at thermal gas-kinetic rates if the {\HH O}/OH ratio 
is of order unity, in mild violation of extant observational limits.
\keywords{ interstellar medium -- molecules }
}
\maketitle

\section {Introduction.}

The roster of polyatomic molecules known to exist in diffuse
interstellar gas has been steadily enlarged over the last
fifteen years and now numbers nearly one dozen.  Two of
these, C$_3$ \citep{MaiLak+01,RouFel+02,OkaTho+03} and
H$_3$\p\ \citep{McCHun+03} were recently discovered in
optical and near-infrared absorption spectra.  The others
have accumulated gradually over the last twenty or so years
in cm-wave and mm-wave radiofrequency absorption data and 
now include \hhco \citep{FedWil82,Nas90,MarMoo+93,LisLuc95a,MooMar95}, 
\hcop \citep{LucLis96}, 
\ammon\ \citep{Nas90}, \cch\ \citep{LucLis00a}, \c3h2\ 
\citep{CoxGue+88,LucLis00a}, HCN and HNC \citep{LisLuc01} and 
$\HH$S and HCS\p \citep{LucLis02}.  


Observationally, the polyatomics observed with radio 
techniques are often seen to be strongly 
and directly related to simpler species observed at optical 
wavelengths, {\it i.e.} \hcop\ to OH \citep{LisLuc96,LucLis96}
HCN and HNC to CN \citep{LisLuc01} $etc.$  These connections point 
to fundamental {\it lacunae} in our understanding of the chemistry of 
even the simplest species in diffuse clouds: {\bf simple} models which 
reproduce the abundances of diatomics, 
however easily, invariably fall orders of 
magnitude short of explaining the related polyatomics.
Conversely, the high abundances of some polyatomics
transform other mysteries; the poorly-understood abundances
of CO and CS in diffuse clouds are readily understandable
given the recombination of the observed columns of \hcop\
\citep{LisLuc00} and HCS\p\ \citep{LucLis02} respectively. 

The work described here rounds out to an even dozen the number
of polyatomics known to exist in diffuse and translucent
clouds. \hocp, the energetically-disfavored \hcop\ isomer
whose 89.4874 GHz J=1-0 transition we sought, was indeed detected
along several of the sightlines studied in the other molecular
species mentioned in the preceding paragraphs.  The \hcop-\hocp\
comparison is especially interesting now because both species have
recently been detected over a wide range of physical conditions
in dense gas: our new spectra reinforce an emerging view
which relates molecular abundances in dense and diffuse PDR.  
Section 2 gives details of the observations, which are discussed 
in Sect. 3 and summarized in Sect. 4.

\section{Observations of \hocp}

The 89.4874 GHz J=1-0 transition of \hocp\ was observed at the Plateau de
Bure at various times during the period 2000-2001 toward the five sources
listed in Table 1.  Profiles were taken with 39.1 kHz (0.13 \kms) channel
separation and 70 kHz resolution and subsequently once hanning smoothed,
yielding spectra with the line/continuum rms values as listed in Table 1.

The column density in the lowest (J=0) level of a simple linear
molecule is related to the integrated optical depth of the J=1-0
transition as

$$ N_0 = {{8.0\times10^{12} \pcc \int\tau_{10} dv}\over
{\mu^2 (1-\exp{(-h\nu_{10}/k\Texc)})}} 
\eqno(1)$$

where $\mu$ is the permanent dipole moment of the molecule, 
$\Texc$ is the excitation temperature of the J=1-0 transition 
and $\nu_{10}$ is its frequency.

With the exception of CO \citep{LisLuc98}, the molecules observed 
at mm-wavelengths
in diffuse gas are well-described by assuming that the rotational 
ladder is thermalized at the temperature of the cosmic microwave
background, in which case the total column density
can be calculated straightforwardly from observations of one
transition.  We take the dipole moments of \hocp\ and \hcop\ as
2.8D and 3.93D, respectively, a slight departure from prior work
in which we used 4.07D for \hcop.  Given these assumptions, one
has N(X)/$\int\tau_{10} dv = 2.146 \times 10^{12} \pcc$ for X=\hocp,
$1.093 \times 10^{12} \pcc$ for X=\hcop\ and 
$1.128 \times 10^{12} \pcc$ for X=H$^{13}$CO\p.

\begin{table}
\caption[]{Background sources observed}
{
\begin{tabular}{lcccc}
\hline
Source&l& b & $\sigma_{\rm{l/c}}^1$& $\int\tau(\hocp) dv$ \\ 
\hline
&&&&\kms \\
\hline
B0355+508 & 150.38\degr  & $-$1.60\degr  & 0.0022 &  0.042(0.005) \\
B0415+379 & 161.68\degr & $-$8.82\degr  & 0.0029  & 0.051(0.004)\\
B0528+134 &  191.37\degr &$-$11.01\degr  & 0.0101  &  $<$0.036 \\
B1730-130 &  12.03\degr &$+10.81$\degr  & 0.0081  &  $<$0.026  \\
B2200+420 &  92.13\degr &$-$10.40\degr  & 0.0041&  0.017(0.006) \\
\hline
\end{tabular}}
\\
$^1$ rms noise in line/continuum ratio at 140 kHz (0.46 \kms) 
resolution \\
\end{table}

\begin{table}
\caption[]{Column densities}
{
\begin{tabular}{lccc}
\hline
Source & N(\hocp) & N(\hcop) & N(OH) \\ 
 & $10^{10}\pcc$& $10^{12}\pcc$ & $10^{13}\pcc$ \\ 
\hline
B0355+508 & 9.01(1.07) & 6.51(0.02) & 14.87(0.67) \\
B0415+379 & 10.9(0.90) & 12.8(0.40)$^2$ &  42.50(0.10)\\
B0528+134 & $<7.7^1$ & 2.28(0.02) &  4.72(0.31)\\
B1730-130 & $< 5.7^1$ & 1.25(0.02) &  1.34(0.08)\\
B2200+420 & 3.7(1.3) & 2.58(0.03) &  5.73(0.21)\\
\hline
\end{tabular}}
\\
$^1$ limits are $2\sigma$ \\
$^2$ Entry is 59$\pm 2$ N(H$^{13}$CO\p) \\
\end{table}

\begin{table}
\caption[]{Column density ratios$^1$}
{
\begin{tabular}{lcc}
\hline
Source & N(\hcop)/N(\hocp) & N(\hocp)/N(OH) \\ 
\hline
B0355+508 & 72(9) & $6.1(0.7)\times10^{-4}$\\
B0415+379 & 117(12)& $2.6(0.3)\times10^{-4}$\\
B0528+134 & $>$30 & $<1.6\times10^{-3}$ \\
B1730-130 & $>$ 18 & $<4.3\times10^{-3}$ \\
B2200+420 & 70(24) & $6.5(2.1)\times10^{-4}$\\
\hline
\end{tabular}}
\\
$^1$ limits are $2\sigma$ \\
\end{table}

\section{Observed and expected abundances of \hocp\ in diffuse gas}

\begin{figure}
\psfig{figure=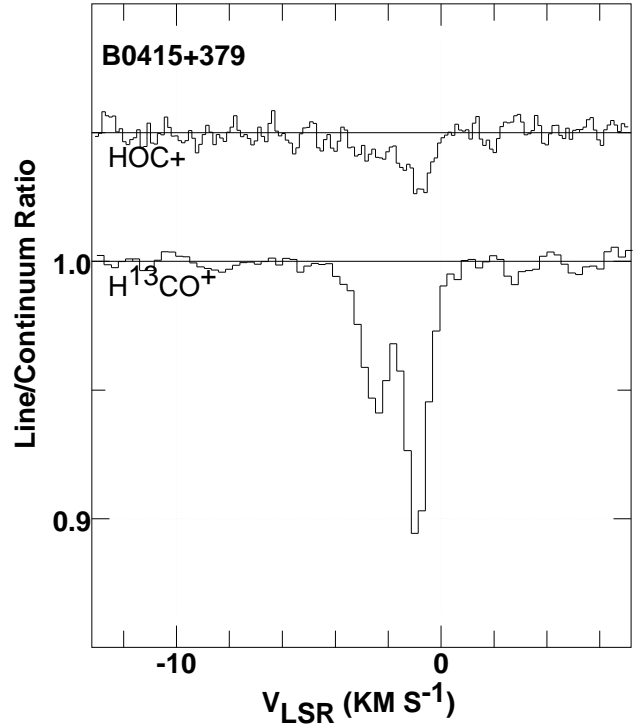,height=9.5cm}
\caption[]{\hocp\ and H$^{13}$CO\p observed toward B0415+379 
(aka 3C111)}\end{figure}

\subsection{Observed abundance of \hocp }

The mm-wave absorption data are summarized in Tables 1-3 and 
line profiles toward two sources are shown in Figures 1 and 2.  The 
\hcop\ and OH column densities are taken from our prior published 
\citep{LucLis96,LisLuc96} and unpublished work; for B0415+379
(see Fig. 1) we assert N(\hcop) = 59$\pm$2 N(H$^{13}$CO\p)
\citep{LucLis98} because of the high optical optical depth
in the main isotope.  \hocp\ is found to have relative abundances 
\hcop:\hocp = 70-120, \hocp:OH $= 3-6\times10^{-4}$ toward
three sources (see Table 3).  The results obtained toward the 
other two, B0528 and B1730, are not of great significance.
The relative abundance of OH, for which it is found that 
X(OH) {\bf = n(OH)/n(\HH)} $ = 1.0\pm0.2\times10^{-7}$ 
along the four sightlines where both N(\HH) and N(OH) have
been measured \citep{LisLuc02}, sets the relative abundance 
scale for \hocp, X(\hocp) $= 3-6\times 10^{-11}$ .

Note that the OH column density toward
B2200+420 is only 20\% larger than that seen along the 
archetypical diffuse line of sight toward \zoph, 
N(OH) = $4.7 \pm 0.7\times~10^{13}~\pcc$ \citep{BlaVan86,Rou96},
while the reddening (\EBV= 0.33 mag toward B2200+420 from \cite{SFD98}) is
the same.  The larger total OH column density seen toward B0355+508 
at low galactic latitude ($-1.6$\degr) is roughly evenly divided among 
5 kinematically-separated diffuse clouds of modest OH and
\hcop\ column density, only two of which have prominent CO
emission \citep{LucLis96,LisLuc96}.  The line of sight toward 
B0415+379 is somewhat darker (see the discussion in 
\cite{LucLis98}) but relatively little of the carbon along 
even this line of sight is in CO and the density is fairly low, 
as gauged by the weakness of emission from species beside CO 
\citep{LucLis96}.

\subsection{Expected abundance of \hocp}

The most obvious path to \hocp\ in diffuse gas is $via$ the reaction 
${\rm C}\p +\HH {\rm O} \rightarrow \hocp + {\rm H}$, which proceeds
with a rate constant $1.8\times10^{-9}~$cm$^3$~\ps\ in the 
UMIST reaction database
\footnote{available at http://www.rate99.co.uk/ }:
these two reactants also produce \hcop\ at half the rate,
but this is not an important source of \hcop.   
We parametrize the unknown abundance of water relative to OH
and assume that all free carbon is once ionized in a fully
molecular gas having a free carbon abundance
N(C)/N(H) = $1.4\times10^{-4}$ \citep{SavSem96}. 
In these terms, the volume formation rate of \hocp\ from the reaction
of C\p\ and water can be expressed as dn(\hocp)/dt
= $5.04\times10^{-20}~$n(\HH)${^2}$ X(\HH {\bf O})/X(OH) cm$^{-3}$ \ps .

\begin{figure}
\psfig{figure=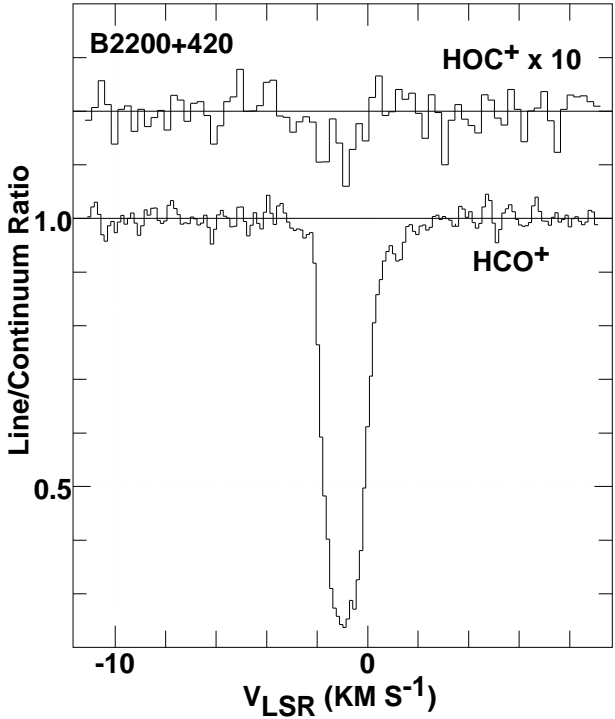,height=9.5cm}
\caption[]{\hocp\ and \hcop\ observed toward B2200+420 
 (aka \bll)}\end{figure}

Both \hcop\ and \hocp\ are formed from the reaction of 
H$_3^+ + {\rm CO} \rightarrow (\hcop~{\rm or}~\hocp) + \HH$.
The \hcop/\hocp\ branching ratio is typically taken 
as 20 \citep{IllJar+82} or more (the current UMIST reaction
database has a branching ratio of 62).  The volume formation rate
from this reaction may be written as dn(\hocp)/dt =
 $2.7\times10^{-23}~$ n(\HH)${^2}$
(X(H$_3^+$) X(CO)/$10^{-12}$) cm$^{-3}$ \ps .
The relative abundance of CO is known along these lines of sight
to be 1-5\% of the free carbon abundance \citep{LisLuc98}, {\it i.e.}
$1-5 \times 10^{-6}$.  The observed amounts of H$_3^+$, while
surprisingly large \citep{McCHun+03} nevertheless still yield 
X(H$_3^+$) $< 10^{-6}$.
Thus the only circumstance in which the reaction of H$_3^+$ and CO
could be competitive in forming \hocp\ would be for very small 
water abundances, in which case neither formation route would 
suffice to form the observed \hocp.

The reaction of CO\p\ and \HH\ forms only \hcop, contrary to the claim
of \cite{FueRod+03} (their Sect. 6.1).

\hocp\ is likely destroyed by two processes which proceed with
very nearly equal rates in diffuse gas. The \hocp\ isomer lies 
some 17,000K higher in energy compared to 
\hcop\ and interconversion by molecular hydrogen
(\hocp\ + \HH $\rightarrow$ \hocp\ + \HH) has recently been
shown to proceed with a rate
constant $ k_I = 4\times 10^{-10}~{\rm cm}^3~$\ps nearly independent of 
temperature between 25K and 300 K \citep{SmiSch+02}.  
This is 70\% smaller than in the UMIST database, enhancing
the prospects for reproducing X(\hocp).
The destruction rate for isomerization can be written 
as dn(\hocp)/dt = ${\bf -4}\times10^{-10}~ n(\HH)$ n(\hocp) 
cm$^{-3}$ \ps.

Dissociative recombination, 
$\hocp + e^- \rightarrow {\rm CO} + {\rm H}$, proceeds
with a rate constant $1.1\times10^{-6}(30/\TK)$ cm $^3$ \ps.
Assuming that free carbon is once-ionized and expressing the 
electron fraction in terms of the ratio of free carbon to 
$\HH$ expected for a fully molecular diffuse gas ({\it i.e.} 
 N(C)/N(\HH) = $2 \times 1.4\times10^{-4}$ \citep{SavSem96}).
The destruction rate from recombination can be written  as 
dn(\hocp)/dt = $-3.1\times10^{-10}~ n(\HH) n(\hocp) (30/\TK)$ 
(X(e)/X(C\p)) cm$^{-3}$ \ps.
The expected electron abundance is a small multiple of the
free carbon abundance \citep{Lis03} so dissociative recombination
and interconversion probably compete quite effectively to destroy
\hocp.

The net destruction rate from both processes is of order
$-10^{-9}$ n(\HH) cm$^{-3}$ \ps\ and the expected abundance of 
\hocp\ is 
X(\hocp) $ = 5 \times 10^{-11}$ X(\HH O)/X(OH) or
X(\hocp)/X(OH) $ = 5 \times 10^{-4}$ X(\HH O)/X(OH).
The observed amounts of \hocp\ can be made in diffuse
gas if the abundance of water is comparable to that of
OH.  Limits on the OH/water ratio in diffuse clouds
N(OH)/N(\HH O) $> 2-3$ were derived by \cite{BlaVan86} toward
\zoph\ and $\zeta$ Per.

Given the weakness of the limits on the water abundance in diffuse 
clouds,  straightforward gas-phase chemistry fails only weakly to 
produce the observed amount of \hocp.  This situation can be 
compared to the case  of \hcop\ where the thermal reaction of 
C\p\ upon the known quantities of OH falls 30-50 times short 
of producing the observed values of N(\hcop) via the reaction 
chain
C$\p + {\rm OH} \rightarrow {\rm CO}^+ + {\rm H}, 
 \HH + {\rm CO}\p \rightarrow \hcop +{\rm H},
\hcop + {\rm e}^- \rightarrow {\rm CO} + {\rm H}$.  
At thermal rates, this series
of reactions (which predicts an \hcop/OH ratio dependent only on 
temperature) produces N(\hcop) scarcely larger than the observed 
N(\hocp).
The predicted abundance of \hcop\ could be increased in quiescent gas 
by hypothesizing that it recombines slowly, but only at the cost
of understanding CO \citep{LisLuc00}.

Clearly, knowledge of the abundance of water is the limiting
factor in assessment of the diffuse cloud chemistry 
of \hocp.

\subsection{Comparison with \hcop\ and \hocp\ in dense dark and bright gas}

The work presented here is the first evidence of \hocp\ in diffuse
gas, but even the case for its existence in the dense
molecular ISM was marginal before the pioneering work of Apponi 
and his collaborators during a brief period in the mid-1990's while
the NRAO 12m telescope was still operating.
To summarize the prior results we note that the 
\hcop/\hocp\ ratio in dense dark gas is
typically as large as 1000-9000 \citep{ZiuApp95,AppZiu97}, but is
found to be much smaller (50-400) in dense PDR like the Orion Bar 
\citep{AppPes+99} and/or NGC7023 \citep{FueRod+03}.  In these PDR
the relative abundance of \hocp\ is X(\hocp) $= 1-3\times10^{-11}$, 
quite similar to the diffuse clouds studied here.

The difference between the \hcop/\hocp\ ratios seen in
diffuse and dark dense gas is largely the result
of the low ionization fraction in the latter.  Whereas \hocp\
is destroyed only slightly faster than \hcop\ in diffuse gas,
the isomerization reaction undergone by \hocp\ is much faster than 
recombination when the electron fraction is $10^{-7}$ instead of
$10^{-4}$.   In dense dark gas \hcop\ and \hcop\ are formed by 
H$_3^+ + {\rm CO} \rightarrow  \HH + (\hcop~{\rm or}~\hocp) $ with
a branching ratio 20-60:1 in favor of \hcop\ and this ratio is 
increased by another
factor of 100 by the unequal destruction rates.
Under favorable conditions, measured \hcop/\hocp\ ratios could 
be good diagnostics of the ionization fraction in dense gas.

Reasoning by analogy from the discussion of diffuse and dark
gas, we infer that the relatively low \hcop/\hocp\ ratios in
dense PDR also reflect the relative weakness of isomerization.
In cases where isomerization is weak both isomers are destroyed
by recombination at roughly equal rates, and their abundance ratio 
is more directly indicative of the relative rates at which the two 
isomers are formed.

\section{Summary}

We have added \hocp\ to the roster of polyatomic molecules known to
exist in diffuse clouds.  The \hcop/\hocp\ ratio, 70-120, is much
smaller than in dense dark clouds but quite comparable to the most
extreme values seen in dense PDR.   The inferred relative abundance
 X(\hocp) $= 3-6 \times 10^{-11}$, similar to that seen in dense
PDR,  can be synthesized at thermal rates 
in quiescent diffuse gas if the water/OH ratio is of order unity.  
This is a mild discrepancy, much less severe than is inferred
for \hcop, as the water/OH ratio is currently {\bf known only} to be less 
than 0.3-0.5.

Several authors have recently noted that hydrocarbons like
\c3h2\ and \cch\ have unexpectedly high abundances in dense PDR 
\citep{FosTey+02,FueRod+03,TeyFos+04}, very much as has been observed in
diffuse clouds \citep{LucLis00a}.  The \hcop/\hocp\ ratios found here
{\bf in diffuse clouds, which are PDR of lower density and presumably 
weaker illumination (hence similar ionization rates per \HH), constitute
another example of chemical similarities between dense and diffuse PDR.}
It will be interesting to see how much more complete a parallel can 
be drawn between the two regimes.

\begin{acknowledgements}

The National Radio Astronomy Observatory is operated by AUI, Inc. under a
cooperative agreement with the US National Science Foundation.  IRAM is
operated by CNRS (France), the MPG (Germany) and the IGN (Spain). 
We owe the staff the Plateau de Bure our thanks for their assistance in 
taking the data.  We thank the referee for helpful comments.

\end{acknowledgements}


\end{document}